\documentclass[12pt,thmsa]{article}
\usepackage{sw20aip}

\input tcilatex
\begin{document}

\title{Hard component of ultra-high energy cosmic rays and vortons}
\author{\and Luis Masperi\thanks{%
On leave of absence from Centro At\'omico Bariloche, S. C. de Bariloche,
Argentina. E-mail: masperi@cbpf.br} \\
Centro Latinoamericano de F\'\i sica \\
Av. Venceslau Br\'az 71 Fundos, 22290 -140 Rio de Janeiro, Brazil \and Milva
Orsaria\thanks{%
Present address: Centro Latinoamericano de Fisica (CLAF), Rio de Janeiro,
Brazil. E-mail: orsaria@cbpf.br} \\
Laboratorio TANDAR\\
Comisi\'on Nacional de Energ\'\i a At\'omica\\
Av. del Libertador 8250, 1429 Buenos Aires, Argentina}
\date{}
\maketitle

\begin{abstract}
Observed events of ultra-high energy cosmic rays may indicate a hard
component for the energy spectrum of their flux, which might have origin in
the decay of long-lived vortons presumably condensed in the galactic halo.
To be consistent with the needed present density, vortons may have been
formed during the breaking of an abelian symmetry contained in a large GUT
group like $E_6$ and a part of them could have survived the destabilization
caused by the electroweak transition.


\end{abstract}

\section{Introduction}

The events of ultra-high energy cosmic rays (UHECR) corresponding to primary
energy above $10^{19}eV$ are difficult to explain\cite{C.T.Norman} with
conventional astrophysical objets, both regarding their acceleration
mechanism and propagation towards the earth due to the interaction with the
cosmic background radiation\cite{K.Greisen} (CBR) if the source is beyond $%
\sim 50Mpc.$

A possible solution of this enigma is given by the so called top-down
mechanism\cite{V.Berezinsky} where long-lived very massive microscopical
objects decay producing the UHECR, which is plausible because so far the
events of the latter appear to be roughly isotropic.

In any case the top-down mechanism would imply physics beyond the standard
model of particles and their interactions (SM). One alternative corresponds
to superheavy relics\cite{J.Ellis} , quasi-stable because their interactions
with the known particles are of gravitational order, which might belong to
the hidden sector where supersymmetry is broken. Another possibility is
given by cosmic strings\cite{C.T.Hill} formed in the phase transition due to
the breaking of a symmetry at the scale of a grand unification theory (GUT).
Though the ordinary Kibble strings\cite{T.W.B.Kibble} consisting of Higgs
and gauge fields might explain\cite{P.Bhattacharjee} the UHECR using their
flux to normalize the model, either the insertion of monopoles forming
necklaces\cite{V.Ber} or the superconducting Witten strings\cite{E.Witten}
with the addition of fermionic fields which give quasi-stable closed loops
called vortons seem more suitable.

It is the purpose of the present work to analyze the details of vortons as
source of UHECR. It had already been seen\cite{L.Masperi} that they may
produce the global flux above $10^{19}eV$ provided their density is
dramatically reduced during the electroweak transition in a way that might
generate the matter-antimatter asymmetry of the universe\cite{M.Orsaria} .

We now describe the energy spectrum of the flux given by the decay of a
superheavy boson emitted by the vorton and conjecture, because of the
hadronization of the resulting quark, that it can constitute the hard
component that might emerge above the GZK cutoff according to the recent
presentation of events\cite{M.Takeda} . Even though vortons should behave as
cold dark matter (CDM) and be concentrated in the galactic halo, we also
study the softening caused by redshift if they were distributed uniformly in
the space as occurs for necklaces and find that it is not possible to
distinguish both cases at present, though in the latter situation a more
important effect coming from the interaction with CBR\ should produce a
depression due to the GZK cutoff followed by a recover of the spectrum\cite
{G.Sigl} caused by its hard nature at emission.

We then face the difficult problem of the dynamics of strings, crucial for
determining the vorton density before and after the electroweak transition.
We assume that the GUT model allows superconductivity to appear at the same
scale of the string formation and evaluate that the delay in the
stabilization of vortons is not too relevant to reduce their density. When
the universe cools down to the electroweak temperature, we estimate the
conditions for the rate of destabilization of vortons due to disappearance
of the Dirac zero-modes which originated the superconductivity of exotic
quarks and the replacement with those of ordinary fermions. The result is
that the collapse of the most abundant short vortons is sufficiently gradual
to avoid the reheating that would dilute the baryogenesis, and the surviving
long ones succeed in absorbing the ordinary fermions with parameters
compatible with the density and lifetime necessary to explain the UHECR.

Finally we discuss which is the possible GUT group consistent with our
mechanism. We see that SO(10) is not adequate since at GUT scale only
vortons with $\nu _R$ might be formed, which could not give the baryogenesis
at the electroweak transition where it is moreover unlikely that long loops
might be stabilized by ordinary fermions. On the contrary, E$_6$ is suitable
for our purposes because at high temperature vortons with exotic quarks may
be formed linked to zero-modes that subsequently disappear at electroweak
scale due to mass effects of the light Higgs, whereas new zero-modes for
ordinary fermions are allowed by the existence of two additional abelian
symmetries of the model apart from the electromagnetic one.

\section{Energy spectrum of UHECR flux from vorton decay}

Considering sources that emit $\stackrel{.}{n}(t)$ UHECR per unit space and
time the total flux on earth will be

\begin{equation}
F=\frac 1{4\pi }\smallint _{t_{in}}^{t_0}dt\text{ }\stackrel{.}{n}(t)\quad
\left( \frac{a(t)}{a(t_0)}\right) ^3\quad ,  \label{e1}
\end{equation}
where $a$ is the scale parameter of universe, $t_0$ its present age and $%
t_{in}$ an initial time which depends on the assumed distribution of sources
but in no case is earlier than that which by redshift produces energies at
least $\sim 10^{19}eV.$

If the sources are quasi-stable objects like vortons with density $n(t)$
each one having at time $t$ a probability $\Gamma $ per unit time of
emitting UHECR observed on earth, 
\begin{equation}
F=\frac 1{4\pi }\smallint _{t_{in}}^{t_0}dt\text{ }n(t)\left( \frac{a(t)}{%
a(t_0)}\right) ^3\Gamma \quad \text{ }.  \label{e2}
\end{equation}

We assume that the vorton emits by tunneling a superheavy particle X, Higgs
or gauge boson of GUT scale, which very quickly decays in quarks and
leptons, the former giving the UHECR by hadronization. We will consider two
cases: the more plausible one in which vortons, behaving as nonrelativistic
particles, condense in the galactic halo, and the other extreme alternative
analogous to ordinary cosmic strings in which they are still uniformly
distributed in space.

\subsection{Condensation in halo}

In this case redshift may be neglected and $\Gamma =\frac{N_c}\tau $ , where 
$\tau $ is the vorton lifetime for emission of X whose decay produces $N_c$
UHECR. Then the total flux will be 
\begin{equation}
F_h=\frac{N_c}{4\pi }n_h(t_0)\frac{\Delta t}\tau \text{ },  \label{e3}
\end{equation}
where $\Delta t\sim 50kpc$ due to the halo size. As it will be seen in the
next section, with $\tau $ larger than $t_0$ and $n_h(t_0)$ a fraction of
the dark matter in the galactic halo, a flux of UHECR of the expected order
is obtained.

We now turn to the energy spectrum $F(E)$ such that $F=\smallint dE$ $F(E),$
where the limits of integration correspond to the UHECR range. From the
probability distribution 
\begin{equation}
\frac{d\Gamma }{dE}=\frac 1\tau \sum_{i=1}^{N_c}\delta (E-E_i)\text{ },
\label{e4}
\end{equation}
the flux spectrum will be 
\begin{equation}
F_h(E)=\frac 1{4\pi }n_h(t_0)\frac{\Delta t}\tau \sum_{i=1}^{N_c}\delta
(E-E_i)\text{ }.  \label{e5}
\end{equation}

To compare with observations, one must average on the intervals $\Delta E_i$
separating neighbouring particles in energy 
\begin{equation}
\overline{F}_h(E_i)=\frac 1{\Delta E_i}\text{ }\frac{n_h(t_0)}{4\pi }\text{ }%
\frac{\Delta t}\tau \text{ }.  \label{e6}
\end{equation}

If events are equally spaced in $\log E,$ which is plausible in
hadronization with QCD except for the upper limit $\sim m_X$ as it will be
discussed below, 
\begin{equation}
\Delta E_i\sim E_i\text{ },\text{ }\overline{F}_h(E_i)\sim \frac 1{E_i}\text{
, }  \label{e7}
\end{equation}
corresponding to a hard component compared with the standard behaviour for
lower energy which is roughly $F(E)\sim E^{-3}.$

\subsection{Uniform distribution in universe}

Since we consider vortons as quasi-stable particles, if we assume a
hypothetical uniform density $n_u(t_0)$ at present, from Eq.(2) the spectrum
at earth would be 
\begin{equation}
F_u(E)=\frac 1{4\pi }n_u(t_0)\int_{t_{eq}}^{t_0}dt\frac{d\Gamma }{dE_{em}}%
\text{ }\frac{dE_{em}}{dE}\text{ },  \label{e8}
\end{equation}
where the lower limit is approximatively the matter-radiation equivalence
time $t_{eq}$ in order to include into UHECR particles redshifted from $\sim
10^{24}eV$ which we take as the maximum energy$.$ If we disregard for the
moment the attenuation due to interaction with CBR, the probability
distribution is the same as Eq.(4) but referred to emission energy\cite
{Y.Chikashige} E$_{em}$%
\begin{equation}
\frac{d\Gamma }{dE_{em}}=\frac 1\tau \sum_{i=1}^{N_c}\delta (E_{em}-E_i)%
\text{ },  \label{e9}
\end{equation}
where the relation with observed energy is given by redshift $z$%
\begin{equation}
E_{em}=(1+z)E\text{ , }1+z=\left( \frac{t_0}t\right) ^{\frac 23}\text{ }.
\label{e10}
\end{equation}

Therefore the spectrum Eq.(8) becomes, being $t_0>>t_{eq}$ $,$%
\begin{equation}
F_u(E)=\frac 3{8\pi }n_u(t_0)\frac{t_0}\tau E^{\frac
12}\sum_{i=1}^{N_c}\frac 1{E_i^{\frac 32}}\theta (E_i-E)\text{ }.
\label{e11}
\end{equation}

The total flux for UHECR defined for $E>E_0$ is, with $E_0<E_i,$%
\begin{equation}
F_u=\int_{E_0}^\infty dE\text{ }F_u(E)=\frac 1{4\pi }n_u(t_0)\frac{t_0}\tau
\left[ N_c-\sum_{i=1}^{N_c}\left( \frac{E_0}{E_i}\right) ^{\frac 32}\right] 
\text{ }.  \label{e12}
\end{equation}

Averaging the spectrum in intervals defined for convenience as 
\begin{equation}
\overline{F}_u(E_j)=\frac 1{\Delta E_j}\int_{E_{j-1}}^{E_j}dE\text{ }F_u(E)%
\text{ },  \label{e13}
\end{equation}
and with the above hypothesis of $\Delta E_j\sim E_j$%
\begin{equation}
\overline{F}_u(E_j)=\frac 1{4\pi }\ n_u(t_0)\ \frac{t_0}\tau \ \frac
1{\Delta E_j}\left[ 1-\left( \frac{E_{j-1}}{E_{N_c}}\right) ^{\frac
32}\right] \quad \text{ },  \label{e14}
\end{equation}
it is clear that redshift will produce a softening of the law $\frac 1{E_j}$
through the bracket factor, which will be slight except for the highest $j$.

\subsection{Comparison with observations}

Considering that the probability per unit time for a vorton to emit
particles with energy between $E_L\simeq 10^{19}eV$ and $E_H\simeq 10^{24}eV$
from the decay of X is 
\begin{equation}
\int_{E_L}^{E_H}\frac{d\Gamma }{dE_{em}}\text{ }dE_{em}=\frac{N_c}\tau \text{
},  \label{e15}
\end{equation}
the definition of the average on intervals according to Eqs. (9) and (7) 
\begin{equation}
\frac{d\Gamma }{dE_{em}}\simeq \frac 1\tau \frac 1{E_{em}}\text{ },
\label{e16}
\end{equation}
gives, from Eq.(15), N$_c\sim 10$ which is a reasonable value compared with
the extrapolation of fragmentation functions\cite{Yu} . This is consistent
with the total energy emitted by a vorton per unit time, being $m_X\sim E_H,$%
\begin{equation}
\int_0^{E_H}dE_{em}\frac{d\Gamma }{dE_{em}}E_{em}=\frac 1\tau E_H\text{ }.
\label{e17}
\end{equation}

Therefore we may take the equally spaced particles in $\log E$ according to $%
E_1\simeq 10^{19}eV,$ $E_2\simeq 10^{19.5}eV\;.\;.\;.\;E_9=10^{23}eV,$ $%
E_{10}\simeq 10^{23.5}eV,$ and $E_0\simeq 10^{18.5}eV,$ so that with the
previous definition $\Delta E_j=\left( 1-\frac 1{10^{0.5}}\right) E_j$ quite
compatible with Eq.(16).

For the vortons condensed in the halo, since from Eq.(6) the flux in each
bin is the same, one might roughly normalize it at the observed value for $%
10^{19}eV$ \textit{i.e.} 
\begin{equation}
\frac{n_h(t_0)}{4\pi }\frac{\Delta t}\tau =\frac 1{km^2\text{ }yr}\text{
\quad .}  \label{e18}
\end{equation}
Being the mass of a vorton $\sim N_Lm_X$ and representing a fraction $f$ of
the halo average energy density $\sim 0.3$ $\frac{GeV}{cm^3}$ it turns out 
\begin{equation}
\frac{n_h(t_0)}{4\pi }\frac{\Delta t}\tau \simeq \frac f{N_L}\frac{t_0}\tau 
\frac{10^7}{km^2\text{ }yr}\text{ },  \label{e19}
\end{equation}
so that for $\tau \sim t_0$ and $N_L\sim 10^3$ which is a sensible number as
will be discussed in Sec.III $,$ a fraction $f$ $\sim 10^{-4}$ of dark
matter would be enough. More precisely, from the recent presentation of data
shown in Fig.1, there seems to be an extragalactic component above the
``ankle'' and then the hard component dominating beyond the GZK cutoff. With
this interpretation the contribution of vortons is a small part of flux at $%
10^{19}eV$ and $f$ may be even two orders of magnitude smaller. The fact
that our fit normalizes the hard component at $\sim 10^{20}eV$ allows to
reproduce data with $m_X\sim 10^{15}GeV$ at variance with a similar
discussion for superheavy relics\cite{M.Birkel} .

If we instead imagine vortons uniformly distributed, they should constitute
a fraction of the critical density of universe $\rho _c(t_0)\simeq
10^{-29}\frac g{cm^3}.$ Disregarding the redshift depression and GZK cutoff $%
J=\frac 1{4\pi }n_u(t_0)\frac{t_0}\tau $ gives a flux 3 times larger than
that of Eq.(19) because the smaller $n_u$ is compensated by the larger $t_0$
compared to $\Delta t.$ The decrease due to redshift corresponds to the
bracket in Eq.(12) which is just 0.98. Similarly the bracket in the energy
spectrum of Eq.(14) gives $F_u(E)\propto \frac 1{E^{(1+K)}}$ where $K\simeq
0.011$ as is seen by the fit of Fig.2. Therefore it would not be possible to
distinguish the two cases with the present statistics by the redshift
effect. But certainly more important is the effect of the GZK cutoff which
is not so drastic for the hard vorton component\cite{G.Sigl} giving an
effective spectrum $F_u(E)\propto \frac 1{E^2}$ between $10^{19}$ and $%
10^{20}eV$ and a recovering behaviour $\propto \frac 1{E^{1.3}}$ in the
range from $10^{20}$ to $10^{21}eV.$ Therefore the fraction $f$ of dark
matter should be one order of magnitude larger than in the case of
condensation in halo. Similarly, it has been evaluated\cite{M.Blanton} that
the hard component of UHECR emitted from observed galaxies avoids the GZK
cutoff.

Regarding the absolute contribution of other galaxies apart from ours, since
luminous matter is $\Omega _L\simeq 5$ x $10^{-3}$ and halos are $10$ times
larger $\Omega _h\simeq 5$ x $10^{-2}$, a fraction of vortons $\sim 10^{-6}$
would give $5$ x $10^{-8}$ of critical density. This would be 200 times
smaller than the required extragalactic vortons as said above. This
estimation of the negligible contribution of halos of other galaxies is
consistent with detailed computations\cite{O.E.Kalashev} .

More statistics is needed to test the anisotropy in favour of larger mass
concentration in the halo case. It is interesting that UHECR above $5$x$%
10^{19}eV$ seems to show\cite{G.Medina} a small anisotropy not related to
the structure of the local universe. One must remark that the uniform
distribution of superheavy relics is critically constrained by the diffuse
gamma flux at GeV scale from EGRET requiring a low extragalactic magnetic
field $\sim 10^{-12}Gauss.$

A hard component similar to Eq.(7) appears from accurate calculation\cite
{M.Birkel} with QCD but it is also suggested by semiquantitative arguments.
On one side the quark which comes from decay of X may be considered as a
ultrarelativistic particle suffering a constant force due to friction.
Therefore the decrease of its momentum is proportional to time and since
this corresponds to hadronization, to produce a more energetic particle it
takes more time,\textit{\ i.e.} the law of Eq.(16) follows. In another way,
the hadronization results from the emission of a gluon with energy similar
to that of the quark and close to its direction. Thus the transition
amplitude is $\sim \frac 1E$ and the number of final states $EdE$ to keep a
fixed angle around the quark,\textit{\ i.e.} the probability of
hadronization from vorton 
\begin{equation}
d\Gamma \simeq \frac 1\tau \frac 1{E\text{ }^2}EdE\text{ },  \label{e20}
\end{equation}
is Eq.(16).

All what said in this Section is valid both for vortons and superheavy
relics. The next one will be devoted to dynamics of vortons to discuss in
which way they may have the required density for UHECR.

\section{String dynamics and vorton densities}

The dynamics which may lead to the vorton density necessary for UHECR
consists of several stages. First of all we will discuss the possible vorton
density above the electroweak (EW) transition temperature. For the formation
of vortons there are four temperatures\cite{B.Carter} : production of
ordinary strings $T_X$, appearance of superconductivity $T_\sigma $,
incorporation of fermionic carriers by loops giving the protovortons $T_f$,
elimination of excess energy relaxing to classically stable configuration $%
T_r.$ We will take the alternative that superconductivity, \textit{i.e. }%
zero-modes of x-y Dirac equation in presence of bosonic string fields,
appears in the same phase transition where strings are generated by Kibble
mechanism \textit{i.e.} $T_\sigma =T_X$ . This allows the possibility of
producing the matter-antimatter asymmetry\cite{M.Orsaria} through the
subsequent elimination of most of vortons. The number density of
protovortons formed at the temperature $T_f$ is\cite{B.Car} 
\begin{equation}
n(T_f)\simeq \frac 1{\left[ \xi (T_f)\right] ^3}\text{ },  \label{e21}
\end{equation}
where $\xi $ is the length below which smaller scale structure will have
been smoothed by friction damping. Once vortons are formed, their density
evolution corresponds to quasi-stable particles in an expanding universe and
is related to that of protovortons which originate them by 
\begin{equation}
n_v(T)=n(T_f)\left( \frac T{T_f}\right) ^3\text{ }.  \label{e22}
\end{equation}
If vortons could be formed\cite{R.Brandenberger} in the friction stage of
Kibble strings produced in a phase transition of GUT scale $T_X$ , $\xi $ is
a sort of average between the string damping time $\tau _d\simeq \frac{T_X^2%
}{T^3}$ and the Hubble time $H^{-1}\simeq \frac{m_{pl}}{T^2}$ i.e. 
\begin{equation}
\xi \simeq \sqrt{\tau _dH^{-1}}=\left( m_{pl}\right) ^{\frac 12}\frac{T_X}{%
T^{\frac 52}}\text{ .}  \label{e23}
\end{equation}
Since we take $T_\sigma =T_X$ , this will be true if the time necessary for
fermions to be absorbed by the string forming protovortons is short.
Moreover the subsequent interval for getting rid of excess of energy to
reach the optimum radius of stabilized vortons must preserve the friction
regime $T>\frac{T_X^2}{m_{pl}}$ to avoid radiation from the not yet static
protovortons. If this occurs, from Eqs.(21-23) 
\begin{equation}
n_v(T)\simeq \left( \frac{T_f}{m_{pl}}\right) ^{\frac 32}\left( \frac{T_fT}{%
T_X}\right) ^3\text{ },  \label{e24}
\end{equation}
and since as we will show $T_f$ is close to $T_X$ the large density 
\begin{equation}
n_v(T)\simeq \left( \frac{T_X}{m_{pl}}\right) ^{\frac 32}T^{\text{ }3}
\label{e25}
\end{equation}
is enough to produce the expected matter-antimatter asymmetry if most of
vortons collapse at the EW transition\cite{M.Orsaria} . In the chiral case
which corresponds to our fermionic carriers, their number in the loop is\cite
{B.Car} 
\begin{equation}
N\approx \xi T_X  \label{e26}
\end{equation}
and from $\xi (T_f)$ of Eq.(23) $N\approx \left( \frac{m_{pl}}{T_X}\right)
^{\frac 12}$ which for the GUT scale $T_X\sim 10^{16}GeV$ will give the most
abundant vortons with $N\sim 10$ carriers.

According to some approches\cite{C.Martins} , it is hard that vortons can be
formed in the friction stage of string dynamics. We will use a simplified
model to estimate $T_f$ and $T_r$ to verify that they are in the friction
regime. Adopting the phenomenological criterium of seeing whether the
fermions can be absorbed by the Kibble string to give way to a protovorton
before the loop collapses, the rate of their incorporation may be given by 
\begin{equation}
\frac{dn_i(t)}{dt}=\alpha \text{ }n_o\quad ,\text{ }  \label{e27}
\end{equation}
where $n_o$ is the outher fermionic density. We assume, subject to
consistency, that the process is fast enough to disregard the universe
expansion and consider the above fixed number of fermions to be incorporated
to the string. $n_o$ will roughly correspond to a radiation mode at GUT
scale $n_o\sim T_X^3.$ We require that the density inside the string passes
from zero at the time of Kibble string formation $t=0$ $,$ to a final value $%
n_i$ in one direction due to field fluctuation such that $n_iL_p\frac
1{T_X^2}\simeq N$ $,$ where $L_p$ is the protovorton length and $\frac
1{T_X} $ its width . The probability per unit time of fermion absorption $%
\alpha =\alpha _0\sqrt{1-v^2}$ will include the Lorentz factor for time
dilatation due to the loop velocity of contraction $v$ which is
qualitatively consistent with the statement that vorton formation is more
difficult for large velocity\cite{C.Martins} . The probability in the rest
frame must be $\alpha _0=h$ $m_X$ because it increases with the difference
between the mass of the fermion outside due to symmetry breaking at GUT
scale and its zero value inside the string which favours its flow there. $h$
is a free parameter presumably smaller than 1 if the mass of the exotic
fermion is smaller than that of the GUT gauge boson as occurs for most of
ordinary fermions compared to EW bosons. Therefore we require from Eq.(27) 
\begin{equation}
\frac N{L_p}=h\text{ }m_X\int_0^{\Delta t}dt\sqrt{1-v^2}\text{ },
\label{e28}
\end{equation}
for $\Delta t$ smaller than the collapse time of string $.$ Since the
protovorton must subsequently lose the excess of energy to reach
stabilization as a vorton of length $L_v=\frac N{m_X},$ the original length
of the string will be $L_0>L_p>L_v$ $.$

For the step between the formation of the Kibble string of length $L_0$ up
to the absorption of N carriers to have a protovorton of length $L_p$ we
consider that the loop contracts due to the tension $\mu =m_X^2$ $,$ which
is also the energy per unit length, according to 
\begin{equation}
-m_X^2=\frac d{dt}\left( E\text{ }v\right) \text{ ,}  \label{e29}
\end{equation}
with $E=L$ $m_X^2,$ $v=\frac{dL}{dt}$ $.$ The result is 
\begin{equation}
\frac L{L_0}=\frac 1{\left( 1+v^2\right) ^{\frac 12}}\text{ , }  \label{e30}
\end{equation}
which, inserted into Eq.(28), gives 
\begin{equation}
\frac 2N=h\text{ }\lambda _0^2\text{ }\frac{L_p}{L_0}\text{ }I\text{ },
\label{e31}
\end{equation}
where $\lambda _0=\frac{L_0}{L_v}$ , $I=\int_0^{\pi -2ar\sin \left( \frac{L_p%
}{L_0}\right) }dz\left( \cos z\right) ^{\frac 12}.$

The procedure is to choose a value of $h$ and $N=10$ as said above if
Eq.(25) holds\cite{R.Brandenberger} . Assuming an initial length $\lambda
_0, $ one obtains $\frac{L_p}{L_0}$ from Eq.(31) and then the protovorton
velocity $v_p$ from Eq.(30) which must be consistent with the subsequent
step to determine $\lambda _0$ .

This corresponds to the delay between protovorton and vorton. When the
former is planar it has an energy 
\begin{equation}
E=\mu L+\frac{N^{\text{ }2}}L\text{ ,}  \label{e32}
\end{equation}
where, to the Kibble contribution of the first term, the kinetic one for
massless fermionic carriers is added. Vortons are the classically stable
loops corresponding to the minimum of $E$.

But on top of the energy $E$ of the ground state of the string Eq.(32), for
the protovorton one may add the energies corresponding to the possibility of
twisting its $N$ pieces. The energy of each twist may be taken as 
\begin{equation}
E_t=e\frac N{L-L_v}\text{ ,}  \label{e33}
\end{equation}
because for $L>>L_v$ it is reasonable that it corresponds to the current, $e$
being the charge. When $L\rightarrow L_v,$ $E_t\rightarrow \infty ,$
indicating that only the plane state of the vorton is possible. As a
consequence, the thermodynamical calculation of minimization of total free
energy $F_{tot}=E+F$ must be done with $F=-T$ $\ln Z$ in terms of the
partition function 
\begin{equation}
Z=\sum_{m=0}^N\left( _m^N\right) e^{-\left( \frac{mE_t}T\right) }\text{ ,}
\label{e34}
\end{equation}
where $m$ is the number of twists. $F_{tot}\left( L\right) $ is a function
flatter than $E(L)$ but with the same minimum. Its variation is given by 
\begin{equation}
-dF_{tot}\left( L\right) =d\left( Ev\right) v+SdT-\widehat{\mu }dL\text{ ,}
\label{e35}
\end{equation}
where $\widehat{\mu }$ is the chemical potential for the protovorton but,
being the last term equivalent to $-dE,$ the equation to be solved involves
in the left hand side only the variation of the partition function of
twists. Since moreover the second term on the right hand side is small
because it corresponds to the reheating produced by the disappearance of a
piece of protovorton transformed into radiation, being at this stage the
density of protovortons much smaller than that of the radiation, Eq.(35)
reduces essentially to 
\begin{equation}
T\text{ }d\ln Z(L)\simeq d\left[ E(L)v\right] v\text{ .}  \label{e36}
\end{equation}

Eq.(36) is solved numerically with the condition that at the end, when the
stabilized vorton is reached, $v=0.$ Taking $e\simeq 0.3$ the partition
function is evaluated for the two contributions $\widetilde{N}$ $E_t<T$ and $%
\left( \widetilde{N}+1\right) E_t>T$ as $Z=Z_1+Z_2$ with 
\[
Z_1=\sum_{m=0}^{\widetilde{N}}\left( _m^N\right) e^{-\left( \frac{mE_t}%
T\right) }\simeq \sum_{m=0}^{\widetilde{N}}\left( _m^N\right) \left( 1-m%
\frac{E_t}T+...\right) ,\text{ }\left( a\right) \text{ } 
\]
\begin{equation}
Z_2=\sum_{m=\widetilde{N}+1}^N\left( _m^N\right) e^{-\left( \frac{mE_t}%
T\right) }\simeq \left( _{\widetilde{N}+1}^{\text{ }N}\right) e^{-\left( 
\widetilde{N}+1\right) \frac{E_t}T}+...\text{ . }\left( b\right) \text{ }
\label{e37}
\end{equation}
Thus Eq.(36) allows to go back step by step obtaining the velocity for each
possible initial protovorton length.

As said before, the end of the absorption of fermions must be joined to the
beginning of the process of stabilization of protovortons. In Fig.3 the two
cases of $h=\frac 15$ and $h=\frac 1{13}$ are shown. From these curves the
total interval of time from the formation of the Kibble string to the birth
of the stabilized vorton may be obtained according to 
\begin{equation}
\Delta t_{tot}=L_v\int_1^{\lambda _0}\frac 1{\left| v\right| }d\left( \frac
L{L_v}\right) \text{ .}  \label{e38}
\end{equation}
The results are respectively $\Delta t_{tot}=3.2$ x $10^{-38}\sec $ and $4.5$
x $10^{-38}\sec $ for the two cases of $h$.

Therefore, since the time after big bang of the GUT transition is at least $%
10^{-36}\sec $, the delay for the formation of vortons appears to be small
so that $T_r\approx T_f\approx T_X$ and their density Eq.(25) should hold.

The next step is to analyze what happens to vortons, whose density evolves
according to Eq.(25), when the universe cools down to the temperature $%
T_{EW} $ for the electroweak transition. If the GUT model is such that the
zero-modes for heavy quarks disappear, most of vortons collapse and the
non-equilibrium process may allow to produce the needed matter-antimatter
asymmetry\cite{M.Orsaria} . But if all vortons lose these zero-modes
instantaneously the reheating would be so large that the baryogenesis would
result very much diluted\cite{W.B.Perkins} .

The rate of destabilization of vortons depends on the model, and presumably
it is due to the mixture of Higgs mechanisms at GUT and EW scales where the
latter is responsible for the loss of zero-modes which allowed the exotic
fermion to be stable inside the string. Therefore it is likely that, being
the rate of decay of the fermion outside the string $\sim \alpha _{GUT}$ $%
m_X,$ the rate of destabilization of vorton is 
\begin{equation}
\gamma \sim \alpha _{GUT}\text{ }m_X\left( \frac{T_{EW}}{T_X}\right) ^2\sim
10^{11}\sec ^{-1}\text{ },  \label{e39}
\end{equation}
which is smaller than the Hubble parameter at the beginning of the EW
transition thus smoothing considerably the reheating effect\cite{M.Orsaria} .

Also depending on the model new zero-modes, this time corresponding to
ordinary fermions, may appear at the EW temperature. Then vortons stabilized
by them may be formed. But since the loops are collapsing due to the loss of
the original zero-modes, one has to see whether the ordinary fermions
succeed in being absorbed to reach the required density before the strings
disappear. It is understandable that the surviving vortons will be the long
ones.

The string is composed now by superheavy bosons X in its inner core and by
the EW bosons in the outer part. The length of the inner core is still $%
\frac N{m_X}$ and its width $\frac 1{m_X}$ so that the energy per unit
length both of the Higgs potential and magnetic GUT contributions is $\sim
m_X^2$ . The outer ring will have a width $\sim \frac 1{m_W}$ and
consequently a length $\frac N{m_W}$ so that the Higgs potential and
magnetic weak contributions will be $\sim m_W^2<<m_X^2$ . Regarding the
density of ordinary fermions to have a new vorton, it must be $n_i\sim m_W^3$
.

An argument similar to the one following Eq.(27), and since outside the
string now $n_o\sim T_{EW}^{\text{ }3},$ indicates that 
\begin{equation}
n_i\simeq h\text{ }T_{EW}^{\text{ }4}\text{ }\Delta t\text{ ,}  \label{e40}
\end{equation}
which requires that $\Delta t$ is smaller than the collapse time $\tau _c$.
Since for masses of ordinary fermions $h\sim 10^{-3}$ and $\tau _c%
\stackunder{\sim }{>}\frac N{m_{EW}}$ because all the string configuration
including its electroweak part must collapse together to avoid energy
divergences, Eq.(40) will be satisfied by long strings with $N\sim 10^3$ and
not for the most abundant vortons with $N\sim 10.$

According to the discussion of Sec.2, the ratio of the density of long
vortons $N_L\sim 1000$ which allow to produce UHECR and that of short ones $%
N\sim 10$ Eq.(25) which would give the expected baryogenesis should be 
\begin{equation}
\frac{n_L(T)}{n_v(T)}\sim 10^{-27}-10^{-28}\text{ }.  \label{e41}
\end{equation}

It is a very delicate matter to explain this ratio which cannot be related
to that of the corresponding Boltzmann factors for vortons but more
reasonably to the Kibble loops at the beginning of the acquisition of
fermions which means that their energy is one half of that of vortons of the
same length L according to Eq.(32), giving a ratio which is still too small.

But one must remember that, at variance from vortons which are plane, the
protovortons which originated them may be rough and therefore have entropy.
Thinking that a protovorton is a chain of $N$ objects that can be horizontal
or vertical, the degeneracy without considering the energy of twists is $%
d\simeq 2^N,$ which can enlarge the density of long vortons in such a way
that Eq.(41) may be satisfied.

The last important point is to assure that the vorton lifetime for emission
of an X is at least of the order of the universe age to allow the production
of UHECR in recent times. The estimation may be done through the bounce
instanton where, to simplify things, we consider the vorton as an object in
the x-y plane so that the Euclidean action $S_E$ may be taken as the
difference of the three-dimensional energy of a tube which contracts
emitting an X and that of the non-contracted one. The most important
contribution to this difference will be given by the gradient of the GUT
Higgs field $\phi $ so that 
\begin{equation}
S_E\sim \int \left( \frac{\partial \phi }{\partial z}\right) ^2\delta \text{ 
}dxdydz\sim \frac{\left( \Delta \phi \right) ^2}{\Delta z}\delta ^2L\text{ }.
\label{e42}
\end{equation}
$\Delta \phi \sim m_X$ because it is the change from broken to unbroken
vacuum.$~\delta ^2\sim \frac 1{m_X^2}$ is the section of the inner core of
the string where GUT fields are concentrated. Finally $\Delta z\geq
m_X^{-1}, $ because the length for the variation of the tube size must be at
least of the order of the Compton length of the emitted X particle. Since $%
L\simeq \frac{N_L}{m_X},$ Eq.(42) with all the above estimations gives $%
S_E\leq N_L$ and the lifetime for the emission of one X is, being the vorton
mass $m_v\sim N_Lm_X,$%
\begin{equation}
\tau \sim \frac 1{m_v}e^{S_E}\leq \frac 1{N_Lm_X}e^{N_L}\text{ },
\label{e43}
\end{equation}
which, as an order of magnitude, ensures $\tau $ larger than $t_0$ for $%
N_L\sim 10^3.$

In this way we have shown the feasibility of the mechanism for UHECR
according to Eq.(19).

\section{GUT models and possible vortons}

To build a Kibble string it is necessary to break an abelian $\widetilde{U}%
(1)$ symmetry different from the electromagnetic $U(1).$ If the model is
simply 
\begin{equation}
U(1)\text{ x }\widetilde{U}(1)\rightarrow U(1)\text{ },  \label{e44}
\end{equation}
one such infinite string is stable. But if $\widetilde{U}(1)$ is contained
in a larger group G, a discrete symmetry $Z_N$ of the continuous $\widetilde{%
U}(1)$ must remain unbroken to avoid that the corresponding monopoles make
the string unstable by cutting it. This depends on the Higgs mechanism and
in general the discrete symmetry does not survive if the Higgs field
corresponds to the fundamental representation of G.

On the other hand the string will become superconducting if a fermion
acquires mass through a Higgs that winds it. Subsequent phase transitions
caused by different Higgs fields may produce the disappearance of a previous
zero-mode.

\subsection{SO(10)\ }

It has 45 generators and one of its maximum subgroups is $SU(5)$ x $%
\widetilde{U}(1)$, suitable for string formation.

There should be four subsequent breakings with the indicated relevant Higgs
multiplets 
\begin{equation}
SO(10)_{\overrightarrow{\text{ }45\text{ }}}SU(5)\text{ x }\widetilde{U}(1)_{%
\overrightarrow{\text{ }126\text{ }}}SU(5)\text{ x }Z_{2\text{ }}
\label{e45}
\end{equation}
\[
_{\overrightarrow{\text{ }45\text{ }}}SU(3)_c\text{ x }SU(2)_L\text{ x }%
U(1)_Y\text{ x }Z_{2\overrightarrow{\text{ }10\text{ }}}SU(3)_c\text{ x }U(1)%
\text{ x }Z_2\text{ }. 
\]

The decomposition in $SU(5)$ x $\widetilde{U}(1)$ of the involved Higgs and
fermion multiplets is 
\begin{eqnarray}
45 &=&1^0+10^2+\overline{10}^{-2}+24^0  \label{e46} \\
126 &=&1^5+\overline{5}^1+10^3+\overline{15}^{-3}+45^{-1}+50^1  \nonumber \\
10 &=&5^1+\overline{5}^{-1}  \nonumber \\
16 &=&\overline{5}^{\frac 32}+10^{-\frac 12}+1^{-\frac 52}\ .  \nonumber
\end{eqnarray}

The first breaking must be done by a $\phi _{45}$ non vanishing in the $1^0$
component to keep the $SU(5)$ x $\widetilde{U}(1)$ invariance. The second by
a $\phi _{126}=\phi _{1^5}$ to preserve $SU(5)$ symmetry. Regarding
fermions, since $16\ $x$\ 16=126+120+10,$ only the second breaking produces
mass for one of them, the $\nu _R$, through a Majorana term which is
invariant under $\widetilde{U}(1)$ as is seen from Eq.(46). There also
stable infinite strings\cite{A.C.Davis} are produced with winding number $%
n=1 $%
\begin{equation}
\phi _{1^5\overrightarrow{\quad r\rightarrow \infty \quad }}\eta _{%
\widetilde{U}}\ e^{i\theta }\qquad ,\quad \widetilde{A}_{\theta 
\overrightarrow{\quad r\rightarrow \infty \quad }}\frac 1{5\widetilde{e}}\
\frac 1r\quad ,  \label{e47}
\end{equation}
and loops will be classically stabilized as vortons by the superconducting
current of $\nu _R$. Since $SU(5)$ is still valid, which will presumably
require SUSY, $\eta _{\widetilde{U}}\sim 10^{16}GeV.$

The third phase transition does not affect the string because it is done by
a $\phi _{45}$ in $24^0,$ with components which break $SU(5)$ but keep the
invariance under $SU(3)_c$ x $SU(2)_L$ x $U(1)_Y$, that does not feel the $%
\widetilde{U}(1)$ charge $\widetilde{e}.$

On the contrary the last breaking, the EW one, has several consequences.
Since ordinary fermions get Dirac mass from the $SU(5)$ products 
\begin{equation}
\overline{5}\ \text{x}\ 10=5+45\qquad ,\quad 10\ \text{x}\ 10=\overline{5}+%
\overline{45}+50\ ,  \label{e48}
\end{equation}
the $5^1$ part of $\phi _{10}$ will give mass to the quark $u$ and $%
\overline{5}^{-1}$ to quark d. Due to the fact that in $SO(10)$ these masses
come from a single mass term $16\ $x$\ 16$, to avoid the degeneracy of $u$
and $d$ either one takes different expectation values in the neutral
components of $5^1$ and $\overline{5}^{-1},$ or we repeat this term twice
with ad-hoc coefficients, one multiplied by $\phi _{10}$ nontrivial in $5^1$
and the other with $\phi _{10}$ nontrivial in $\overline{5}^{-1}$ with the
same expectation value.

Regarding the string, $\phi _{10}$ cannot be constant everywhere because the 
$\widetilde{U}(1)$ charge would give a contribution to the covariant
derivative $D_\theta $ producing a divergent energy. Therefore we must
accept the possibility of winding $\phi _{5^1}$ and $\phi _{\overline{5}%
^{-1}}$ and also the inclusion of a neutral $Z$ contribution 
\begin{eqnarray}
D_\theta \phi _{5^1} &=&\left( \frac 1r\frac \partial {\partial \theta }-i%
\widetilde{e}\widetilde{A}_\theta -ig_\varphi Z_\theta \right) \phi _{5^1}%
\text{ },\text{ }  \label{e49} \\
D_\theta \phi _{\overline{5}^{-1}} &=&\left( \frac 1r\frac \partial
{\partial \theta }+i\widetilde{e}\widetilde{A}_\theta +ig_\varphi Z_\theta
\right) \phi _{\overline{5}^{-1}}\text{ . }  \nonumber
\end{eqnarray}
If we assume that the winding number of $\phi _{5^1}$ is $m$, the condition
to cancel its covariant derivative is 
\begin{equation}
m-\frac 15-\chi =0\ ,  \label{e50}
\end{equation}
where $\chi $ is the contribution of $Z.$ Obviously, to cancel also the
covariant derivative of $\phi _{\overline{5}^{-1}}$ , which behaves as $\phi
_{5^1}^{*}$ , the same condition Eq.(50) holds meaning that it would wind
with $-m$.

But the minimization of the magnetic energy of $Z$ requires $m=0$ so that $%
\phi _{5^1}$ and $\phi _{\overline{5}^{-1}}$ do not wind and no zero-modes
appear for ordinary fermions.

Moreover $\phi _{10}$ gives rise to a coupling $\overline{\nu }_R\nu _L$
and, since $\phi _{10}$ does not wind, this small massive contribution to
the state of $\nu _R$ makes the corresponding zero-mode disappear so that
all the vortons would collapse at the EW phase transition\cite{S.C.Davis} .

It must be noticed that the situation would change\cite{E.Witten} if an
additional Higgs, which does not generate mass, $210=5^{-4}+\overline{5}%
^4+....$ with expectation value in $5^{-4}$ is present. To compensate the
covariant derivative of such a nonwinding field 
\begin{equation}
D_\theta \phi _{5^{-4}}=\left( \frac 1r\frac \partial {\partial \theta }+i4%
\widetilde{e}\widetilde{A}_\theta -ig_\varphi Z_\theta \right) \phi
_{5^{-4}}\quad ,  \label{e51}
\end{equation}
$Z_\theta $ needs to behave $Z_\theta \rightarrow \frac 45\ \frac
1{g_\varphi }\ \frac 1r$ which introduced into Eq.(49) requires $m=1$. In
this rather artificial way, ordinary fermions would have zero-modes and GUT
vortons should not collapse.

\subsection{ E$_6$}

It has 78 generators and one maximum subgroup is $SO(10)$ x $\overline{U}(1)$
with the relevant chain of breakings 
\begin{eqnarray}
&&E_{6~\overrightarrow{\text{ \quad }78\quad \text{ }}}SO(10)\ \text{x}\ 
\overline{U}(1)_{\overrightarrow{\text{ \quad }27\quad \text{ }}}SO(10)....
\label{e52} \\
&&...SU(3)_c\ \text{x}\ SU(2)_L\ \text{x}\ U(1)_{Y~\overrightarrow{\quad 
\text{ }27\quad \text{ }}}SU(3)_c\ \text{x\ }U(1)\quad .  \nonumber
\end{eqnarray}

The descompositions in $SO(10)\ $x$\ \overline{U}(1)$ of the fundamental and
adjoint representations are respectively 
\begin{eqnarray}
27 &=&1^1+10^{-\frac 12}+16^{\frac 14}  \label{e53} \\
78 &=&1^0+16^{-\frac 34}+\overline{16}^{\frac 34}+45^0\ .  \nonumber
\end{eqnarray}

The first breaking must be done by $\phi _{78}\equiv \phi _{1^0}$ to keep
the invariance of $SO(10)\ $x$\ \overline{U}(1)$ and since $27\ $x$\ 27=%
\overline{27}_S+351_S+351_A$ , no fermion gets mass in it. Fermion masses
come from Higgs in $27$ or $351^{*}$ (which explains the decomposition of $%
126$ in Eq.(46)). Thus the second breaking done by $\phi _{27}\equiv \phi
_{1^1}$ to preserve the invariance of $SO(10)$ , gives mass to the exotic
fermions contained in $10^{-\frac 12}$ of $27$ through a term $\phi
_{1^1}\psi _{10^{-\frac 12}}\psi _{10^{-\frac 12}}$ . These exotic fermions,
one quark $D$ of charge $-\frac 13$ with $3$ colours and an electron $E$
with its neutrino $N$ may be the carriers of a vorton based on the string 
\begin{equation}
\phi _{1^1}\rightarrow \eta _{\overline{U}}\ e^{i\theta }\qquad ,\quad 
\overline{A}_\theta \rightarrow \frac 1{\overline{e}}\ \frac 1r\quad .
\label{e54}
\end{equation}

Due to the fact that 27 is the fundamental (spinorial) representation, a $%
Z_2 $ symmetry is not preserved and the infinite string is not absolutely
stable. However it is stable enough if the scale $\eta _{\overline{U}}\sim
10^{16}GeV$ for the breaking of $\overline{U}(1)$ is at least one order of
magnitude lower than that of breaking of $E_6.$

In the last breaking, an expectation value of Higgs in $10^{-\frac 12}$
gives mass to ordinary fermions through a term $\phi _{10^{-\frac 12}}\psi
_{16^{\frac 14}}\psi _{16^{\frac 14}}$ . Regarding the influence of $\phi
_{10^{-\frac 12}}$ on the string, the difference with the case of $SO(10)$
is that now the $\overline{U}(1)$ charge $-\frac 12\overline{e}$ is common
to both $5^1$ and $\overline{5}^{-1}$ of $SU(5)$ x $\widetilde{U}(1),$ 
\textit{i.e.} 
\begin{equation}
D_\theta \phi _{5^1}=\left( \frac 1r\frac \partial {\partial \theta }+i\frac{%
\overline{e}}2\overline{A}_\theta -i\widetilde{e}\widetilde{A}_\theta
-ig_\varphi Z_\theta \right) \phi _{5^1}\text{ },\text{ }\left( a\right)
\label{e55}
\end{equation}
\[
D_\theta \phi _{\overline{5}^{-1}}=\left( \frac 1r\frac \partial {\partial
\theta }+i\frac{\overline{e}}2\overline{A}_\theta +i\widetilde{e}\widetilde{A%
}_\theta +ig_\varphi Z_\theta \right) \phi _{\overline{5}^{-1}}\text{ . }%
\left( b\right) 
\]
If $\phi _{5^1}$ winds at the EW scale as $\phi _{5^1}\rightarrow \eta _{EW}$
$e^{im\theta },$ to avoid energy divergence coming from Eq.(55a) it must be 
\begin{equation}
m+\frac 12-\chi =0\text{ ,}  \label{e56}
\end{equation}
where now $\chi $ is the contribution from $\widetilde{A}_\theta $ and $%
Z_\theta .$

If the minimization of energy favours $\chi =-\frac 12,$ from Eq.(56) $m=-1$
and since $\varphi _{5^1}$ gives mass to $u$ and $\nu $ they will have
zero-modes and the corresponding carriers will run in the direction $-z$ of
the string axis. Now from Eq.(55b) with $\chi =-\frac 12$ the energy
divergence is avoided if $\phi _{\overline{5}^{-1}}$ does not wind and,
since it gives mass to $d$ and electron, they will have no zero-modes.
Conversely if energy minimization favours $\chi =\frac 12,$ $\phi _{%
\overline{5}^{-1}}$ will wind as $\phi _{\overline{5}^{-1}}\rightarrow \eta
_{EW}$ $e^{im\theta }$ with $m=-1,$ $d$ and $e$ will have zero-modes and
they will be the vorton carriers running along $-z$ of string axis.

Therefore, with a scheme based on $E_6$, vortons formed at scale $\eta _{%
\overline{U}}$ with exotic fermions E and D would remain essentially stable
down to scale $\eta _{EW}$ where would incorporate new carriers either $u$
and $\nu $ or $d$ and $e.$

But if the EW breaking is done with a Higgs in 27 of $E_6$ which has not
only expectation value in its component $10^{-\frac 12}$ but also in $%
16^{\frac 14},$ there may be an ordinary-exotic mass term\cite{J.L.Rosner} $%
\phi _{16^{\frac 14}}\psi _{10^{-\frac 12}}\psi _{16^{\frac 14}}$ , which
will mix E and D with $e$ and $d$. Being $10=5^1+\overline{5}^{-1},$ $5^1$
contains $D,$ $\overline{E}$ and $\overline{N}$ and $\overline{5}^{-1}$ $%
\overline{D},$ $E$ and $N.$

Therefore if the situation that minimizes energy is the first described
above, i.e. $\chi =-\frac 12$ , $d$ and $e$ will have no zero-modes and
their mixing with D and E will destroy the zero-modes of the latter
producing our baryogenesis mechanism. On the other hand, the new zero-modes
of $u$ and $\nu $ would be responsible for the quasi-stability of the long
vortons from the EW age to our days giving rise, through quantum decay, to a
possible source of UHECR.

This is in fact what happens. To cancel the large distance covariant
derivative of $\phi _{10^{-\frac 12}}$ Eq.(55) avoiding the divergence of
energy, one needs, together with the already existing $\overline{A}_\theta $
of Eq.(54), 
\begin{equation}
\text{ }\widetilde{A}_\theta \rightarrow \frac{c_1}{\widetilde{e}}\text{ }%
\frac 1r\text{ , }Z_\theta \rightarrow \frac{c_2}{g_\varphi }\text{ }\frac 1r%
\text{ ,}  \label{e57}
\end{equation}
with $c_1+c_2=\mp \frac 12$ which would allow to choose $\left| c_1\right|
=\left| c_2\right| =\frac 14$ in both cases to minimize the magnetic energy $%
\frac 12\left( \widetilde{B}^2+B_Z^2\right) .$

But we now have the additional condition of cancelling the covariant
derivative of $\phi _{16^{\frac 14}}$ . Because of the decomposition of 16
Eq.(46) in representations of $SU(5)$ x $\widetilde{U}(1)$ , the mixing term 
$\overline{D}$ $d$ may come from $\overline{5}^{-1}$ x $10^{-\frac 12}$ and,
due to the fact that $\overline{5}$ x $10=5+45$ , invariance is obtained
multiplying by a $\phi _{16^{\frac 14}}$ with expectation value in $%
\overline{5}^{\frac 32}$ \textit{i.e.} $\phi _{\overline{5}^{\frac 32}}\psi
_{\overline{5}^{-1}}\psi _{10^{-\frac 12}}$. To respect the symmetry $%
SU(3)_c $ x $U(1)$ the nonvanishing component of $\phi _{\overline{5}^{\frac
32}}$ must be the neutral one so that its covariant derivative is 
\begin{equation}
D_\theta \phi _{\overline{5}^{\frac 32}}=\left( \frac 1r\frac \partial
{\partial \theta }-i\frac{\overline{e}}4\overline{A}_\theta -i\frac 32%
\widetilde{e}\widetilde{A}_\theta -ig_\varphi ^{^{\prime }}Z_\theta \right)
\phi _{\overline{5}^{\frac 32}}\quad .\text{ }  \label{e58}
\end{equation}
Assuming a behaviour $\phi _{\overline{5}^{\frac 32}}\rightarrow \eta _{EW}$ 
$e^{im\theta }$ $,$ the cancellation of energy divergence requires 
\begin{equation}
m-\frac 14-\frac 32c_1+c_2=0\quad ,  \label{e59}
\end{equation}
because $g_\varphi ^{^{\prime }}=-g_\varphi $ as $g_\varphi $ corresponds to 
$5$ and $g_\varphi ^{^{\prime }}$ to $\overline{5}$.

With the condition appropriate for our model $c_1+c_2=-\frac 12$ , Eq.(59)
is satisfied with $c_1=-0.3$ $,$ $c_2=-0.2$ and $m=0$ producing less
magnetic energy than that of the alternative case $c_1+c_2=\frac 12$ which
would require much more difference between $\left| c_1\right| $ and $\left|
c_2\right| .$

The other mixing term $\overline{d}~D$ comes from $\overline{5}^{\frac 32}$
x $5^1$ times a nonvanishing $1^{-\frac 52}$ of $\phi _{16^{\frac 14}}$ with
a covariant derivative 
\begin{equation}
D_\theta \phi _{1^{-\frac 52}}=\left( \frac 1r\frac \partial {\partial
\theta }-i\frac 14\overline{e}\ A_\theta +i\frac 52\widetilde{e}\widetilde{A}%
_\theta \right) \phi _{1^{-\frac 52}}\text{ },\text{ }  \label{e60}
\end{equation}
where $Z$ does not appear because $\phi _{1^{-\frac 52}}$ has no
interactions of $SU(5)$ and therefore of the SM. If $\phi _{1^{-\frac 52}}$
winds with $m$ , to cancel Eq.(60) 
\begin{equation}
m-\frac 14+\frac 52\ c_1=0\quad .  \label{e61}
\end{equation}
and it is easy to see that the case $\chi =-\frac 12$ gives again more
balanced values of $c_1$ and $c_2$ than for $\chi =+\frac 12$ , but less
than those emerging from Eq.(59). Then, since it is not possible to cancel
simultaneously Eqs.(58) and (60) energy saving suggests that only $\phi _{%
\overline{5}^{\frac 32}}$ will be nonvanishing.

For the model of vortons based on $E_6$ it is better if the breaking of $%
SO(10)$ in the missing part of the chain of Eq.(52) does not follow Eq.(45)
but the alternative\cite{P.Langacker} maximum subgroup $SU(4)\ $x$\ SU(2)_L\ 
$x$\ SU(2)_R$ to avoid the formation of strings $\widetilde{U}(1)$ which
would complicate things and also to make the GUT unification avoiding $%
SU(5). $

\section{Conclusions}

If the explanation of UHECR requires the top-down mechanism, this will imply
new physics beyond the SM. In the case of the superheavy quasi-stable
particles, possibly indications on the hidden sector of supersymmetry
breaking will be given. As for the explanation through cosmic strings GUT
will be explored and, particularly with vortons, details of the groups and
Higgs breakings may be revealed.

We have shown the feasibility of this last mechanism. Obviously, due to the
semiquantitative evaluation of several effects, only orders of magnitude
could be given. More accurate analysis must be performed particularly
regarding the quantum decay of vortons and precise details of the
replacement of exotic fermions by ordinary ones at the EW transition as
carriers of superconducting current inside the string, as it has been proved
to be favourable in the case of the $E_6$ symmetry.

Since the features of the UHECR coming from the two quoted alternatives of
top-down mechanism are similar due to the fact that they are based on quark
hadronization, a deep theoretical study of the models will be useful to
falsify some of them.

As for the evidence of sources based on superheavy microscopical objects,
particles or vortons, there are analysis\cite{V.Berezin} to see if there is
a possible anisotropy of observed events towards the massive concentration
in the galaxy. To determine this, as well as the suggested hard component%
\cite{A.M.Hillas} of the spectrum beyond the GZK cutoff and the required
abundance of primary gammas and neutrinos, a much larger statistics is
needed as will be supplied by the Auger project\cite{Pierre Auger} .

\begin{center}
$\mathbf{Acknowledgments}$
\end{center}

LM thanks CONICET, project PICT 0358, and MO CLAF and Fundaci\'on Antorchas,
project A-13798/1, for partial financial supports.

\end{document}